\documentclass[11pt]{article}
\usepackage{moriond,epsfig}
\usepackage{multirow}
\usepackage{bm}
\usepackage{slashed}

\newcommand{\met}{$\slashed{E}_{T}$}
\newcommand{\dzero}{D\O}

\begin{document}
\vspace*{24mm}
\title{DISCOVERY OF SINGLE TOP QUARK PRODUCTION}

\author{Dag Gillberg
  \\(On behalf of the \dzero{} and CDF Collaborations)\\[1mm]}

\address{
  Simon Fraser University, Canada\\
  now Carleton University, Canada, dgillber@physics.carleton.ca
}
\maketitle
\begin{figure}[h]
  \begin{center}
    \psfig{figure=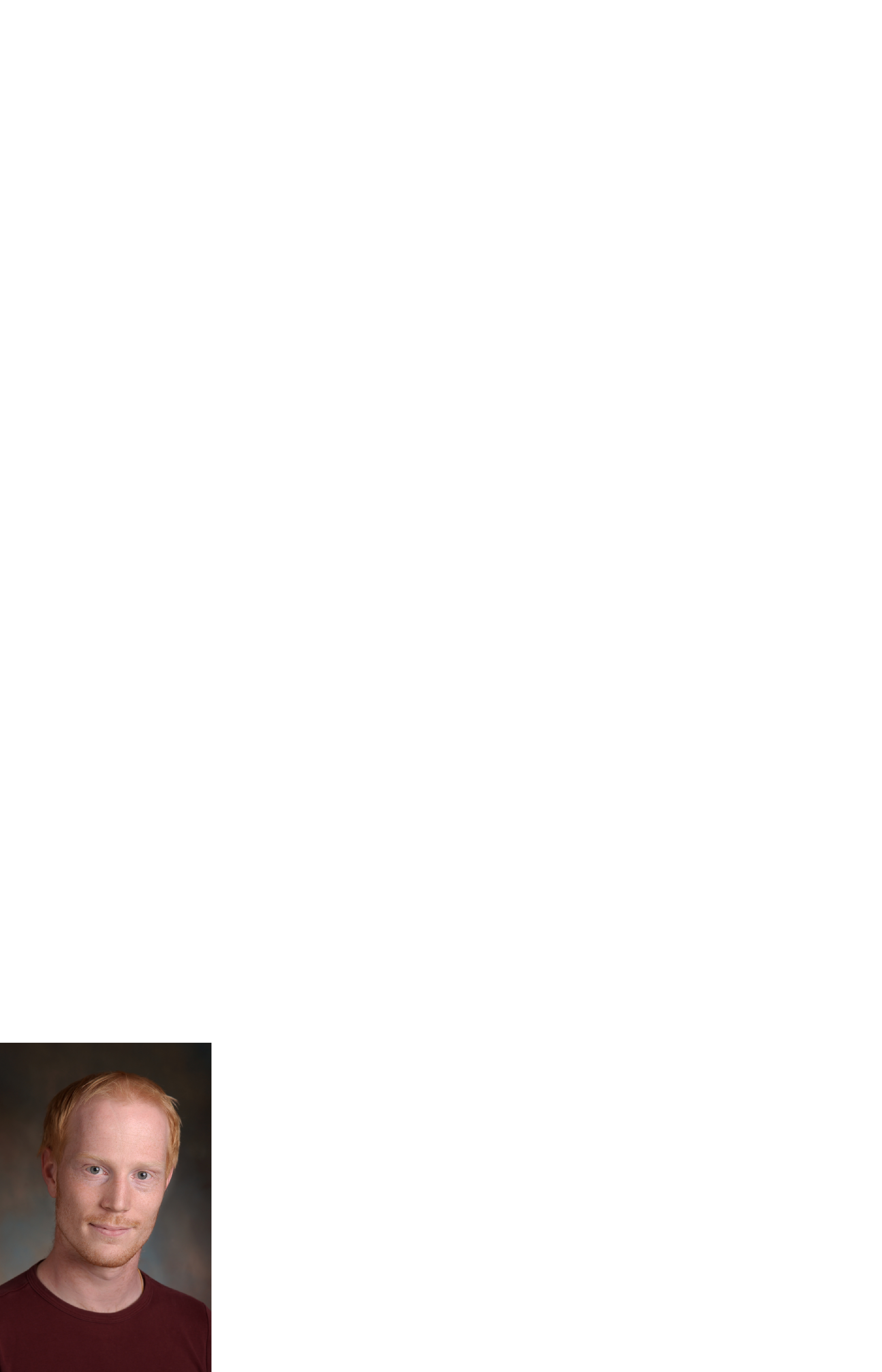,height=35mm}
  \end{center}
\end{figure}
\abstracts{
  The first observation of electroweak single top quark production was
  recently reported by the the \dzero{} and CDF collaborations based on 
  2.3 and 3.2~fb$^{-1}$ of $p\bar{p}$ collision data 
  collected at $\sqrt{s}=1.96$~TeV from the Fermilab Tevatron 
  collider.\cite{d0-prl,cdf-prl} 
  Several multivariate techniques are used to separate the single top
  signal from backgrounds, and both collaborations present
  measurements of the single top cross section and the CKM matrix
  element $|V_{tb}|$. 
}

\section{Introduction}
\label{sec:intro}
\begin{figure}[!h!tbp]
  \centerline{\epsfxsize=3.8in\epsfbox{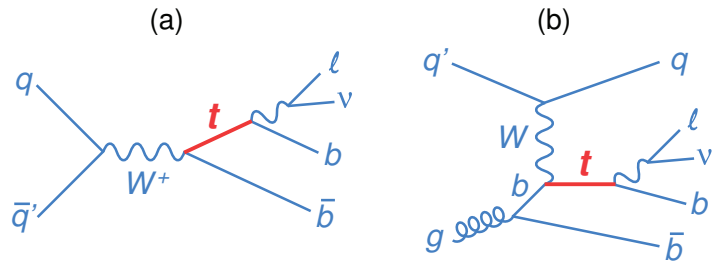}}
  \vspace*{-2mm}
  \caption{
    Representative Feynman diagrams for (a) $s$-channel and (b)
    $t$-channel single top production showing the top quark decays of
    interest.
  }
  \label{fig-feynman}
\end{figure}

The top quark was discovered at the Fermilab Tevatron in
1995~\cite{top-obs-1995} and is the heaviest elementary
particle found so far. At the Tevatron, top quarks are
predominantly produced in pairs via the strong interaction,
but can also be produced singly via an electroweak $Wtb$ vertex.
The two main single top production modes are
the $s$-channel ($tb$) and the $t$-channel ($tqb$) processes 
illustrated in Figure~\ref{fig-feynman}. 
The analyses presented herein assume ${\cal B}(t{\to}Wb)=1$,
and that the $s$-~and $t$-channel modes are produced in the
standard model (SM) ratio.
CDF uses $m_{\rm top}=175$~GeV throughout the analysis, for which the
NLO SM single top cross section $\sigma_{s+t}$ is
2.86~pb.\thinspace\cite{xsec-175} 
\dzero{}~use~$m_{\rm top}=170$~GeV for which $\sigma_{s+t}=3.46$~pb at
(N)NLO.\thinspace\cite{xsec-170}


\section{Motivation}
\label{sec-motivation}
Studies of single top quark events will provide access to properties of the
$Wtb$ coupling,\thinspace\cite{singletop-wtb-heinson}
such as the CKM matrix element $|V_{tb}|$ (see Section~\ref{sec-vtb}).
Single top quark production is also a very important background for
SM Higgs production, 
and studies of the single top final state offer potential for several
discoveries beyond the SM.\thinspace\cite{singletop-newphys} 
For instance, the existence of a new heavy boson (like a charged Higgs)
would enhance the single top $s$-channel cross section while the
existence of flavour-changing neutral currents could enhance the
$t$-channel cross section.

\section{Analysis Strategy}
\label{sec-analysis}

The analyses start by selecting
events where the $W$~boson from the top decays leptonically.
Due to the small single top cross section and the large background
(mainly from $W+$jets processes and $t\bar{t}$), a simple cut-based counting
experiment is not sufficient to verify the presence of single top.
Instead, the strategy used is to apply a fairly loose event selection and
employ sophisticated multivariate techniques to extract the single top
signal from the large backgrounds.

\section{Dataset and Event Selection}
\label{sec-event}

The \dzero{} collaboration 
uses a larger dataset and a looser event selection compared to the previous
analysis,\thinspace\cite{d0-evidence} which reported the first evidence for
single top quark production. The 2.3~fb$^{-1}$ dataset was collected with
the \dzero{} detector using a logical OR of many trigger conditions.
The main event selection criteria applied are an
isolated electron or muon with $p_T>15$~GeV,
\met{}$>20$~GeV, 2-4 jets with $p_T>15$~GeV out of which one jet
has $p_T>25$~GeV and at least one jet must be tagged with a neural
network based $b$-tagging algorithm. Additional 
selection criteria remove multijet background events with
misidentified leptons. 
The numbers of predicted events and the numbers of events observed in
data 
are presented in Table~\ref{tab:yields}.

The CDF collaboration constructs a 3.2~fb$^{-1}$ $\ell$+jets dataset
using a similar, but slightly more stringent, event selection 
than the one used by \dzero{}.
Events are required to have an isolated electron or muon with
$p_T>20$~GeV, \met{}$>25$~GeV and 2-3 jets with $p_T>20$~GeV.
CDF also constructs a \met{}+jets dataset that accepts events where
the $W$~boson decays to a $\tau$ lepton, and those in which the electron
or the muon fail the lepton identification criteria. Events with any
isolated lepton are rejected, and events are required to have
\met{}$>50$~GeV, one jet with $p_T>35$~GeV, a second jet with
$p_T>25$~GeV, and $\Delta R>1.0$ between the two leading jets.
Additional requirements are applied to reduce the large 
instrumental background.\thinspace\cite{cdf-prl}
The number of events observed in the data and
predicted by the modelled background and SM signal are presented in
Table~\ref{tab:yields}.

\begin{table}[!h!tbp]
  \vspace*{-2mm}
  \caption{
    Predicted and observed numbers of events in the \dzero{} and CDF datasets
    for single top ($tb$+$tqb$), the different background components
    and the data. $W$+jets is the largest background and is split into
    events with heavy flavour jets ($W$+HF) and events with fake $b$-jets
    (mistags) in the right table below.
  }
  \vspace*{2mm}
  \hspace{0.01\textwidth}
  \begin{minipage}{0.38\textwidth}
    \centering
    
    \begin{tabular}{lc}\hline\hline
      \textbf{D\O{}}    & $\ell$+jets\\
      ${\cal L}~[\mathrm{fb}^{-1}]$  & 2.3 \\ \hline
      $tb+tqb$    &         223 $\pm$  30 \\
      $W$+jets    &        2647 $\pm$ 241 \\
      $t\bar{t}$  &        1142 $\pm$ 168 \\
      multijet    &         300 $\pm$  52 \\
      $Z$+jets, dibosons &  340 $\pm$  61 \\\hline
      Total prediction &   4652 $\pm$ 352 \\\hline
      Observed    &   4519   \\\hline\hline
    \end{tabular}

  \end{minipage}
  \hspace{0.02\textwidth}
  \begin{minipage}{0.52\textwidth}
    \centering
    
    \begin{tabular}{lcc}\hline\hline
  \textbf{CDF}  & $\ell$+jets & \met{}+jets\\
  ${\cal L}~[\mathrm{fb}^{-1}]$  & 3.2 & 2.1 \\\hline
  $tb+tqb$           &  191 $\pm$   28 &  64 $\pm$  10 \\
  $W$+HF             & 1551 $\pm$  472 & 304 $\pm$ 116 \\
  $t\bar{t}$         &  686 $\pm$   99 & 185 $\pm$  30 \\
  mistags, multijet  &  778 $\pm$  104 & 679 $\pm$  28 \\
  $Z$+jets, dibosons &  171 $\pm$   15 & 171 $\pm$  54 \\\hline
  Total prediction   & 3377 $\pm$  505 & 1404 $\pm$ 172 \\\hline
  Observed           & 3315            & 1411 \\\hline\hline
    \end{tabular}
    
  \end{minipage}
  \label{tab:yields}
\end{table}

\section{Signal-Background Separation}
\label{sec-sig-bkg-separation}
After the event selection, the expected signal is smaller than the
uncertainty on the background (see Table~\ref{tab:yields}). 
Both collaborations use several different multivariate techniques to
further improve the discrimination against the background. 
Each such multivariate technique constructs a powerful discriminant
variable that is proportional to the probability of an event being
signal. The discriminant distribution is used as input to the 
cross section measurement. 
Several validation tests are conducted by studying 
the discriminant output distributions 
in background enriched control samples.



\begin{figure}[!h!tbp]
  \vspace*{-2mm}
  \begin{center}
    \psfig{figure=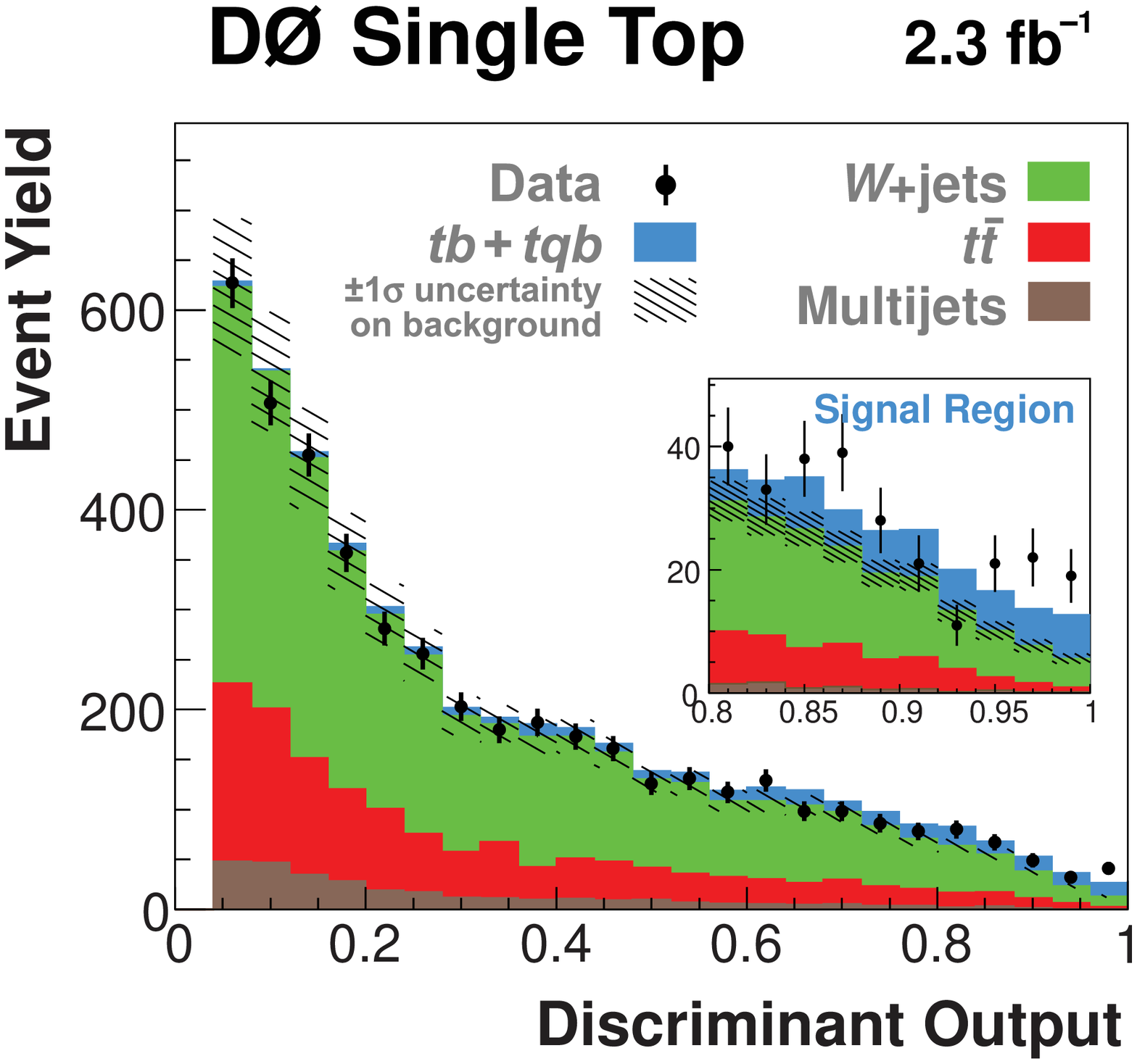,height=60mm}
    \psfig{figure=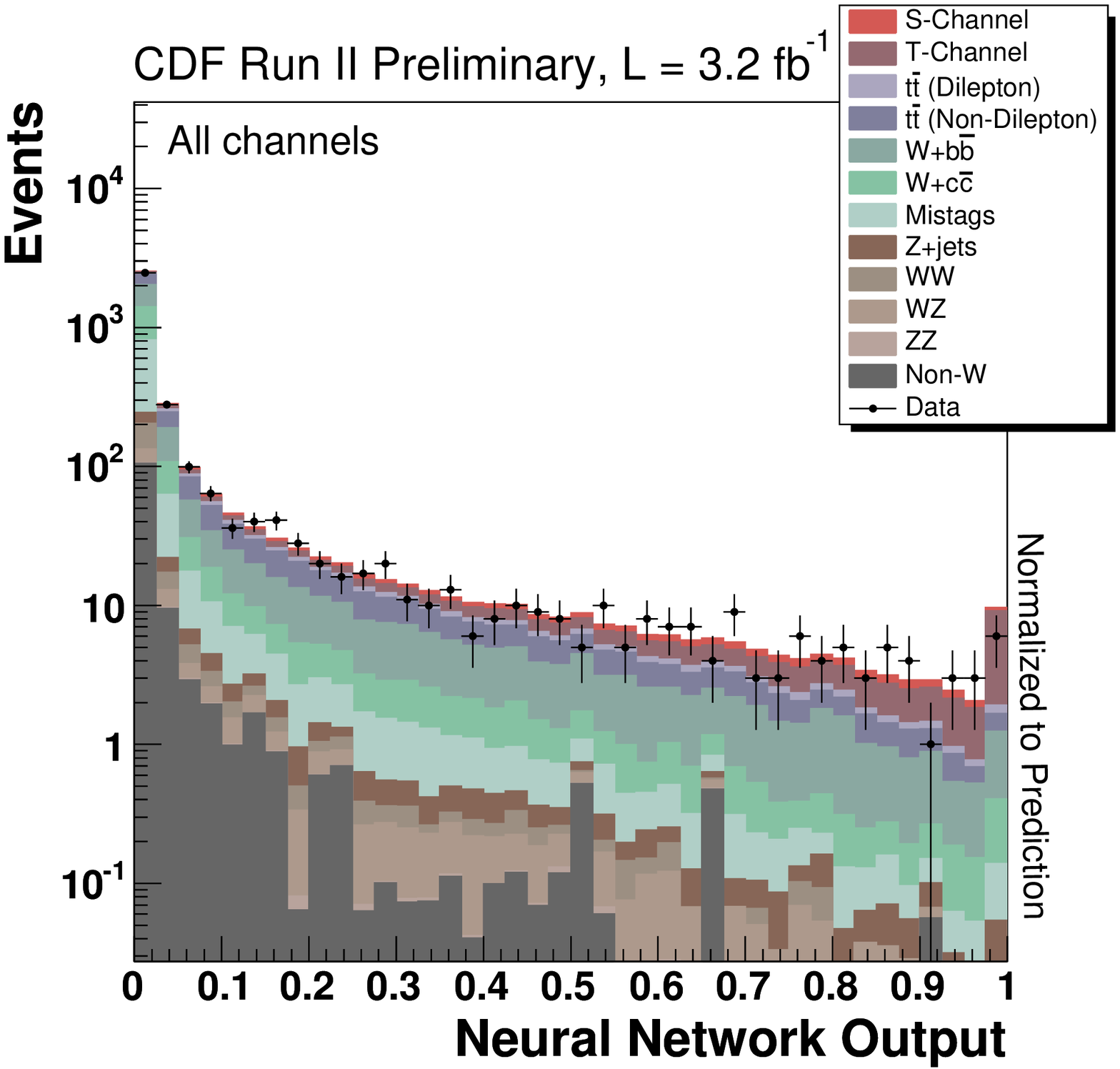,height=60mm}
  \end{center}
  \vspace*{-4mm}
  \caption{
    Combined super discriminants for \dzero{} (left) and CDF (right).
  }
  \label{fig-discriminants}
\end{figure}

The \dzero{} collaboration uses three individual techniques to separate
single top quark events from the background, namely boosted decision
trees (BDT), matrix element method (ME) and Bayesian neural networks
(BNN). 
The BDTs use 64 well-modelled input variables and classify events
based on the outcome of a set of binary cuts. Boosting
is used to further improve the performance.
The ME method calculates a discriminant by relating the 
reconstructed four momenta in the event with the expected
parton-level kinematics. 
The BNN is an average of several hundred neural networks, and is
constructed using 18-28 input variables.
The three multivariate techniques are combined 
separately for each analysis channel
using a second layer of Bayesian neural networks. 
The combined discriminant (Figure~\ref{fig-discriminants}) 
gives a higher expected significance 
than any of the three individual discriminant methods on their own.

The CDF collaboration analyses their $\ell$+jets dataset using five
different multivariate techniques: BDTs, neural networks, the ME
method and two separate likelihood functions. The BDTs
use 20 input variables, the NN analysis constructs four separate NNs from 11-18
input variables, and the ME analysis 
calculates the signal probability from Feynman diagrams. 
Two projective likelihood functions are
constructed using 7-10 input variables; the LF discriminant 
is optimized to find single top quark events in the whole dataset just like 
the other analyses, while the LFS discriminant is 
optimized to be 
sensitive to the $s$-channel process for events with two $b$-tagged jets.
The five $\ell$+jets analyses are combined using a neural network
trained with neuro-evolution,\thinspace\cite{NEAT} which is tuned to maximize
the expected significance. 
The combined discriminant is
shown in Figure~\ref{fig-discriminants}.
CDF also constructs an MJ discriminant from the \met{}+jets dataset
using neural networks.

\vspace*{-6mm}

\section{Results}
\label{sec-results}
Both experiments measure the single top quark production cross section 
from the
discriminant output distributions using a Bayesian binned likelihood
technique.\thinspace\cite{bayes-limits} The statistical and all
systematic uncertainties and
their correlations are considered in these calculations.
The measurements for each individual analysis and for all
analyses combined are presented in Table~\ref{tab:results}. 

\dzero{} estimates the significance of the measurements from a large
ensemble of pseudo experiments (PE) generated from 
background only with all uncertainties included.
The cross section is measured for each PE, and the
expected and observed $p$-values are calculated as the fraction of 
PEs giving a cross section equal to or higher than the SM
$\sigma_{s+t}$ and the measurement in data respectively.
CDF also generates large ensembles of PEs
and obtain the expected and observed $p$-values 
by computing the log-likelihood ratio test statistic~\cite{cdf-prl}
for each PE.
The $p$-values are converted into a number of Gaussian standard
deviations, and are presented in Table~\ref{tab:results}.

\begin{table}[!h!tbp]
  \caption{
    Integrated luminosities, 
    expected and observed significances stated in standard deviations $\sigma$ 
    and measured single top cross sections for the D\O{}
    (left) and CDF (right) analyses. $[\dagger]$~LFS measures $\sigma_{s}$, not $\sigma_{s+t}$.
    \vspace*{2mm}
  }
  \begin{minipage}{0.5\textwidth}
    
  \begin{tabular}{lcccc}\hline\hline
    D\O{} & ${\cal L}$ & \multicolumn{2}{c}{Significance} & Measured \\
    Analysis & {\footnotesize $[ \mathrm{fb}^{-1} ]$} & 
    Exp. & Obs. & $\sigma_{s+t}$ {\footnotesize $[ \mathrm{pb} ]$}
    \\ \hline
    ~ & ~ & ~ & ~ & ~\\[-3.8mm]
    BDT   & 2.3 & $4.3\sigma$ & $4.6\sigma$ & $3.74^{+0.95}_{-0.79}$ \\
    BNN   & 2.3 & $4.1\sigma$ & $5.4\sigma$ & $4.70^{+1.18}_{-0.93}$ \\
    ME    & 2.3 & $4.1\sigma$ & $4.9\sigma$ & $4.30^{+0.99}_{-1.20}$ \\ \hline
    ~ & ~ & ~ & ~ & ~\\[-3.8mm]
    Combined & 2.3 & $4.5\sigma$ & $5.0\sigma$ & $3.94^{+0.88}_{-0.88}$ \\ \hline \hline
  \end{tabular}
  \end{minipage}
  \begin{minipage}{0.5\textwidth}
  \begin{tabular}{lcccc}\hline\hline
    CDF & ${\cal L}$ & \multicolumn{2}{c}{Significance} & Measured \\
    Analysis & {\footnotesize $[ \mathrm{fb}^{-1} ]$} & 
    Exp. & Obs. & $\sigma_{s+t}$ {\footnotesize $[ \mathrm{pb} ]$}
    \\ \hline
    ~ & ~ & ~ & ~ & ~\\[-3.8mm]
    BDT   & 3.2 & $5.2\sigma$ & $3.5\sigma$ & $2.1^{+0.7}_{-0.6}$ \\
    NN    & 3.2 & $5.2\sigma$ & $3.5\sigma$ & $1.8^{+0.6}_{-0.6}$ \\
    ME    & 3.2 & $4.9\sigma$ & $4.3\sigma$ & $2.5^{+0.7}_{-0.6}$ \\
    LF    & 3.2 & $4.0\sigma$ & $2.4\sigma$ & $1.6^{+0.8}_{-0.7}$ \\
    LFS   & 3.2 & $1.1\sigma$ & $2.0\sigma$ & ~~$1.5^{+0.9}_{-0.8}$\thinspace$^{[\dagger]}$ \\
    MJ    & 2.1 & $1.4\sigma$ & $2.1\sigma$ & $4.9^{+2.5}_{-2.2}$ \\ \hline
    ~ & ~ & ~ & ~ & ~\\[-3.8mm]
    Combined & 3.2 & $>5.9\sigma$ & $5.0\sigma$ & $2.3^{+0.6}_{-0.5}$ \\ \hline \hline
  \end{tabular}

  \end{minipage}
  \label{tab:results}
\end{table}

\vspace*{-7mm}

\section{Measurement of $|V_{tb}|$}
\label{sec-vtb}
Both \dzero{} and CDF use their cross section measurements to extract
$|V_{tb}|$. This is possible since the single top cross section is
directly proportional to $|V_{tb}|^{2}$. 
Under the assumptions stated in Section~\ref{sec:intro}, 
CDF measures $|V_{tb}|=0.91\pm
0.11$(stat+sys)$\pm 0.07$(theory)
and sets the limit $|V_{tb}|>0.71$ at 95\%~CL. 
Assuming $m_{\rm top}=170$~GeV (Section~\ref{sec:intro}), 
\dzero{} extracts the limit $|V_{tb}|>0.78$ at 95\%~CL, and 
measures $|V_{tb}f_1^L|=1.07\pm 0.12$, where $f_1^L$ is the strength of the
left-handed $Wtb$ coupling. 

\section{Summary}
\label{sec-summary}
The \dzero{} and CDF collaborations have performed precise measurements of
the 
electroweak single top quark production cross section and the CKM
matrix element $|V_{tb}|$ using 2.3~fb$^{-1}$ and 3.2~fb$^{-1}$ of
data respectively. Both collaborations individually observe a 5.0
standard deviation excess over background in their data and thereby
establish the discovery of single top quark production.

\end{document}